\documentclass[12pt]{iopart}
\usepackage{iopams}
\usepackage{amstext}
\usepackage{setstack}
\usepackage{graphicx}
\usepackage{dcolumn}
\usepackage{bm}
\usepackage{hyperref} 
\usepackage{braket}
\usepackage{color}
\usepackage{framed}
\DeclareGraphicsExtensions{.pdf,.png}
\usepackage[bottom = 3 cm]{geometry}
\usepackage{multirow}
\usepackage{epstopdf}
\usepackage[numbers,sort&compress]{natbib} 

\newcommand{\figref}[1]{Fig.~\ref{#1}}

\newcommand{\secref}[1]{Sec.~\ref{#1}}

\providecommand{\openone}{\leavevmode\hbox{\small1\kern-3.8pt\normalsize1}}

\begin{document}
\title{Implementations for Device-Independent Quantum Key Distribution}

\author{Alejandro M\'{a}ttar$^1$ and Antonio Ac\'{i}n$^{1,2}$}

\address{$^1$ ICFO-Institut de Ciencies Fotoniques, The Barcelona Institute of Science and Technology, 08860 Castelldefels (Barcelona), Spain}
\address{$^2$ ICREA-Instituci\'o Catalana de Recerca i Estudis Avan\c cats, Lluis Companys 23, 08010 Barcelona, Spain}

\begin{abstract}
Device-independent quantum key distribution (DIQKD) generates a secret key among two parties in a provably secure way without making assumptions about the internal working of the devices used in the protocol. The main challenge for a DIQKD physical implementation is that the data observed among the two parties must violate a Bell inequality without fair-sampling, since otherwise the observed correlations can be faked with classical resources and security can no longer be guaranteed. In spite of the advances recently made to achieve higher detecion efficiencies in Bell experiments, DIQKD remains experimentally difficult at long distances due to the exponential increase of loss in the channel separating the two parties. Here we describe and analyze plausible solutions to overcome the crucial problem of channel loss in the frame of DIQKD physical implementations.
\end{abstract}

\section{Introduction}
\label{sec.1}

Quantum information theory has fundamentally changed the way in which we nowadays understand information processing, as many information tasks that are impossible in classical information theory become possible when exploiting intrinsic quantum properties with no classical counterpart. In particular, Quantum Key Distribution (QKD) \cite{Bennett84,Ekert91}, that is, the distribution of a secret key among two honest parties whose security is guaranteed by the laws of quantum physics, has changed and improved the way we understand security: QKD security is no longer based on assumptions on the eavesdropper's computational power, but on the fact that her actions must obey the laws of quantum physics. QKD is arguably the most mature quantum information technology today; in fact, QKD systems are already commercially available these days.

A few years ago, however, several publications reported the successful hacking of some QKD commercial products \cite{Zhao2008, Lydersen2010, Gerhardt2011}. While at first sight this may seem in contradiction with the previous claims, this is not the case as the hackers did not attack the principle but the implementation. The hacking attacks did not imply that the mathematical proofs of QKD security were wrong, or that quantum physics was not valid any longer. What the attacks did imply is that the validity of quantum physics per se is not sufficient to guarantee QKD security. In fact, QKD security proofs are built on several assumptions about the states and measurements used in the protocol that are crucial, but very hard to meet in practice. The mismatch between theoretical security proofs and practical implementations was the loophole that created weaknesses in the protocol and was exploited by the quantum hackers.

Of course, one possible solution to fix mismatches between theory and implementation is to improve the experimental conditions in order to guarantee that the states and measurements required in the protocol are correctly implemented. But it is unclear whether a complete solution to this problem would ever be possible, due to the unavoidable presence of noise and imperfections in any experiment. A completely different solution was proposed in \cite{acin2007}, under the name of device-independent quantum key distribution (DIQKD). Based on previous works on self-testing \cite{mayers1998} and non-signalling quantum key distribution \cite{barrett2005}, the idea was to design QKD protocols whose performance is independent of the internal working of the devices used in the implementation. Security should be guaranteed from the observed statistics without any reference to the states and measurements used to obtain it.

Experimentally, however, DIQKD is hard because the correlations observed by the two parties must violate a Bell inequality \cite{bell1964,BellReview} without post-selection on the data, since otherwise the violation can be faked with classical resources \footnote{This was, in fact, the loophole exploited by the quantum hackers.}. More precisely, and in spite of the advances recently made to achieve higher photo-detecion efficiencies to close the detection loophole in optical Bell experiments \cite{Lynden2015,Giustina2015}, DIQKD remains experimentally difficult at long distances due to the exponential increase of loss in the channel separating the two parties.

Among several attempts to overcome the crucial problem of channel loss in the frame of DIQKD implementations, here we analyze two plausible solutions which have been developed in the recent past. The first solution requires an auxiliary measurement to locally herald the arrival of the system to the user's location. In this way, at each round, the measurement is only chosen provided that the auxiliary measurement -a quantum non-demolition (QND) measurement- succeeded. The second solution is based on the success of a distant measurement performed by some third party -somewhere in between the two main parties- to validate each round; at the end of the protocol, all rounds for which the third party's measurement failed to give the desired outcome can be safely discarded without opening any loophole.

In \secref{sec.2} we provide a general overview on DIQKD, while in \secref{sec.3} we review experimental loopholes. In \secref{sec.4} and \secref{sec.5} we review with detail the two plausible ideas previously mentioned to circumvent the problem of channel loss in DIQKD.  Discussion and conclusions are presented in \secref{sec.6} .

\section{DIQKD}
\label{sec.2}
DIQKD is based on a relaxation of the security assumptions made in standard QKD. In this sense, it follows the line of work of a series of cryptographic protocols designed to be secure against more and more powerful eavesdroppers \cite{mayers1998,barrett2005}. The device-independent scenario \cite{acin2007} is based on a minimal set of fundamental assumptions (see for instance \cite{Pironio2009}): \textit{(i)} Alice and Bob are located in two secure laboratories from which no unwanted classical information can leak out. \textit{(ii)} They have a trusted random number generator, possibly quantum, producing a classical random output. \textit{(iii)} They have trusted classical memories and computing devices and they share an authenticated, but otherwise public, classical channel.  \textit{(iv)} Quantum physics is correct \footnote{This last assumption can even be relaxed if one considers an eavesdropper with supra-quantum power (\textit{e.g} a non-signaling eavesdropper \cite{barrett2005}).}.

From a practical standpoint, DIQKD fixes the main drawbacks of usual QKD. Usual QKD security proofs rely on several assumptions about the quantum systems that are very hard to meet in practice and which compromise the security of these protocols, as recent hacking attacks have shown \cite{Zhao2008, Lydersen2010, Gerhardt2011}. A possible, but challenging, solution to these problems would be to characterize very precisely the quantum devices and try to adapt the security proof to the actual implementation of the protocol. The concept of DIQKD \cite{acin2007,masanes2011,pironio2012,Vazirani2012,MillerShi2014}, on the other hand, applies through its full generality in a minimalist way to these situations as it allows one to ignore all implementation details.

\begin{figure}[h!]
  \begin{minipage}[c]{0.5\textwidth}
    \includegraphics[width=1.3\textwidth]{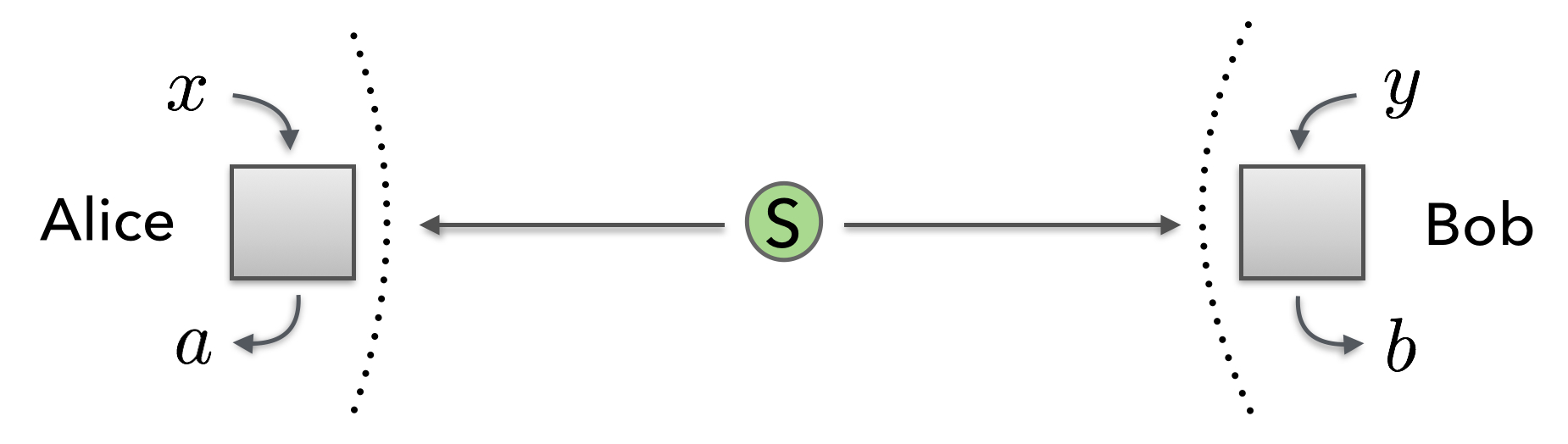}
  \end{minipage}\hfill
  \begin{minipage}[c]{0.5\textwidth}
    \caption{
       \textbf{DIQKD standard protocol.} Alice and Bob receive a quantum system from a source S onto which they perform measurements $x$ and $y$ and retrieve outcomes a and $b$, respectively.
    } \label{fig.1}
  \end{minipage}
\end{figure}

For concreteness, we recall the protocol for DIQKD assessed in Refs. \cite{acin2007,masanes2011,pironio2012}. At each round of the experiment, Alice and Bob receive some unknown quantum state $\rho_{\text{AB}}$ from a source S and perform on it one out of $m_A$ ($m_B$) possible measurements $x=0,1,..,m_A-1$ ($y=0,1,...,m_B-1$)  and retrieve one out of $o_A$ ($o_B$) possible outcomes $a=0,1,...,o_A-1$ ($b=0,1,..,o_B-1$). No other assumption on $\rho_{\text{AB}}$ is made, other than the fact that it is a quantum state. In fact, $\rho_{\text{AB}}$ could be an operator of any dimension, and could even be correlated with another quantum system in the possession of a malicious eavesdropper Eve. The measurements made by Alice and Bob on $\rho_{\text{AB}}$ remain also uncharacterized; the only restriction set is that they shall obey Born's rule in order to reproduce the observed statistics, namely:
\begin{equation}
P(a,b|x,y) = \text{Tr} \left[ \rho_{AB}\ M_{a|x} \otimes M_{b|y} \right].
\label{eq.Born}
\end{equation}
where $\{ M_{a|x}\}$ and $\{ M_{b|y}\}$ are the positive operators defining Alice and Bob's measurements, that is, $M_{a|x}\geq 0$ and $\sum_a M_{a|x}=1$ for all $x$, and similar relations hold for Bob's measurements.

First, Alice and Bob use an authenticated public channel to compare a sample of their data in order to estimate the conditional probability distribution $P(a,b|x,y)$. If $P(a,b|x,y)$ violates a Bell inequality \cite{bell1964,BellReview} $\textbf{g}=\{g_{abxy}\}$  by a sufficient amount, \textit{i.e}, if
\begin{equation}
\sum_{abxy}g_{abxy}P(ab|xy) > g_{loc},
\label{eq.Bell}
\end{equation}
where $g_{loc}$ is the local bound of the inequality, then Alice and Bob can use standard error correction and privacy amplification to distill a secret key out of the remaining data from measurements $x_{raw}=0$ and $y_{raw}=0$.

Security proofs in DIQKD are made from the following observation: under reasonable assumptions, the restrictions set on the observed correlations $P(a,b|x,y)$ by \eref{eq.Born} and \eref{eq.Bell} are sufficient to bound Eve's information on the string of bits that Alice and Bob observe. In \cite{pironio2012}, the only assumption required to prove security is that the quantum memory of the eavesdropper is bounded, which is reasonable given the current status of quantum technologies. It is worth mentioning that a noise-tolerant proof of security without any additional assumption has been given in~\cite{Vazirani2012}. A different and more robust proof has later been derived in~\cite{MillerShi2014}.

\section{Loopholes}
\label{sec.3}

Since Alice and Bob are each located in a secure place and control the information going in and out of their locations (dotted lines in \figref{fig.1}), the value of the inputs $x$ and $y$ and of the outputs $a$ and $b$ does not leak out unwillingly of Alice's and Bob's secure place. Thus, DIQKD is not affected by locality loopholes \cite{BellReview}. In fact, the locality loophole only becomes an issue if one cannot guarantee that the information on the choice of input is not transmitted from one device to the other. But if the devices can leak unwanted information to break the protocol, why shouldn't they simply broadcast, for instance to the eavesdropper, the outputs used to construct the secret key? In our view, making a distinction between inputs and outputs is rather arbitrary and artificial in this context\footnote{This of course does not mean that the locality loophole is not relevant in other contexts in which Bell inequalities are tested.}. Among several other loopholes affecting Bell experiments \cite{Larsson2014}, DIQKD is principally difficult experimentally because of the detection loophole \cite{Pearle1970,Eberhard93}\, which sets a critical overall detection efficiency $\eta$ \footnote{Roughly speaking $\eta$ is the product of the efficiencies of all the physical processes (transmission, coupling, detection, ...) occurring between the source and the users.} below which security can no longer be guaranteed.

It is worth mentioning that very recenly, three independent experiments have managed to successfully produce a loophole-free Bell test; first with light-matter interaction based systems \cite{Hansen2015} and subsequently with purely optical ones \cite{Giustina2015,Lynden2015}. Interestingly, these loophole-free Bell experiments directly enable another device-independent task based on the observation of nonlocal correlations and known as Device-Independent Random Number Generation (DIRNG) \cite{Pironio2010,Pironio2014}. DIRNG relies on the same security assumptions  than DIQKD for the two boxes used in the protocol (see \secref{sec.2}), but with the difference that the two boxes are held by a single party -the one willing to generate random numbers- in the case of DIRNG. Thus, contrary to DIQKD, experimental setups for DIRNG do not need to deal with the problem of long distances.

Indeed, in spite of the technological advances recently made to achieve higher detecion efficiencies in Bell experiments, DIQKD remains experimentally difficult at long distances due to the exponential decrease of transmission efficiency in the channel separating the two parties. In fact, in the standard protocol (see \figref{fig.1}), with polarization entangled pairs of photons used as information carriers and with current optic fibre technology, the detection loophole is already opened for a distance of the order of $\approx 4\ km$ \cite{Gerhardt2011b}.

It is worth noting that partial solutions -namely, semi-device independent approaches- to the problem of DIQKD have been developed in the recent years. One first relaxation to this problem is that of Measurement-Device-Independent Quantum Key Distribution (MDIQKD) \cite{Lo2012}, in which the two parties willing to share a secret key prepare specific quantum states that are sent to a measurement station in between them. The state preparation is device-dependent, as the protocol relies on the preparation of specific quantum states for each round, but the measurement process in the middle remains, indeed, device-independent. Very recently, MDIQKD has been experimentally demonstrated at high rates with continuous variable systems \cite{Pirandola2015}. Another possibility for relaxing DIQKD consists of assuming that one of the two parties trusts his measurement apparatus while the other party remains untrusted. This relaxation is often known as One-Sided-DIQKD \cite{Branciard2012}. In this situation, which is formally based on the observation Einstein-Podolsky-Rosen (EPR) steering correlations \cite{Wiseman2007,Skrzypczyk2014}, the trusted party receives unnormalized quantum states that are conditioned on the observation of an input-output pair of classical labels by the untrusted party. Finally, one last semi-device-independent approach worth to mention consists on assuming that the dimension of the states prepared in the protocol is always bounded. Recently, security proofs have been derived under such ``bounded dimension'' assumption for different prepare-and-measure QKD protocols \cite{Pawlowski2011,Woodhead2015}. These three relaxations discussed here are certainly easier to implement than DIQKD and provide higher key rates, but the price to pay at the level of security seems to be high, since the corresponding setups remain vulnerable to hacking attacks on the internal working of some of the devices used, unless one has a tomographic control of \textit{everything that happens} at the trusted sites, which seems rather unrealistic, as discussed in \secref{sec.1} 

\section{Local heralding}
\label{sec.4}
\begin{figure}[h!]
  \begin{minipage}[c]{0.5\textwidth}
    \includegraphics[width=1.3\textwidth]{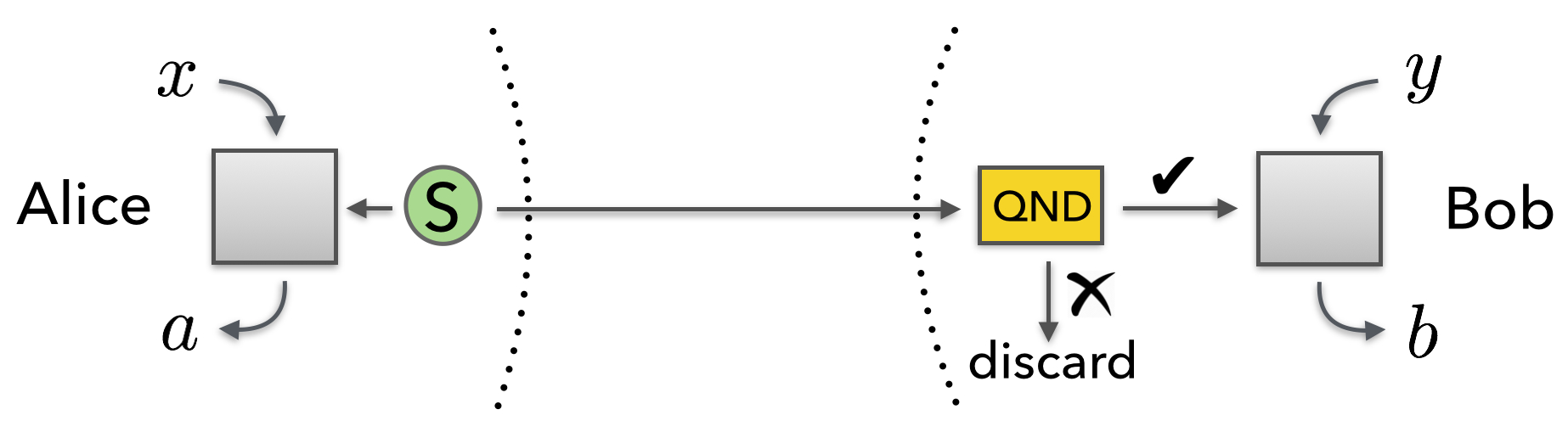}
  \end{minipage}\hfill
  \begin{minipage}[c]{0.5\textwidth}
    \caption{
       \textbf{DIQKD with local heralding.} The source is located within Alice's site. Bob performs an auxiliary QND measurement on his system before choosing his setting $y$.
    } \label{fig.2}
  \end{minipage}
\end{figure}

A first solution to overcome the problem of channel loss in DIQKD was realized by Gisin, Pironio and Sangouard \cite{Gisin2010}. The scheme is based on the heralded noiseless qubit amplification \cite{ralph2009}, which given a state with a vacuum and single-photon component $\alpha\ket{0}+\beta\ket{1}$, it amplifies with some non-zero probability the single-photon component $\alpha\ket{0}+G\beta\ket{1}$, up to normalisation, with $G>1$. The success probability is smaller for higher values of the gain factor $G$. Then idea of~\cite{Gisin2010} is to use the heralded amplification as an approximation to a quantum non-demolition (QND) measurement at Bob's site to herald to him the arrival of his photon, without destroying its carried information (see \figref{fig.2}). In this way, the source can now be placed next to Alice (to avoid loss on her side of the channel), and Bob only measures his system whenever the amplification measurement succeeds.

 The source in Ref. \cite{Gisin2010} is assumed to produce pairs of polarization entangled photons. To perform the noise-less amplification, at each round Bob inserts two single photons with orthogonal polarizations on a beam-splitter of transmittance $T\approx 1$. The reflected modes are jointly measured with his input mode via a Bell state measurement (this defines the amplification measurement). Whenever two clicks are observed in two of the detectors corresponding to orthogonal polarizations at the Bell state measurement, the transmitted mode (output mode) is projected to the original input state. Heralded qubit amplification has been observed in a proof-of-principle experiment recently, first with polarization entangled pairs of photons in the visible regime \cite{Kocsis2013}, and subsequently in the telecom regime with time-bin entangled pairs of photons \cite{Bruno2015}, although not without the presence of the detection loophole. This represents an important achievement for the future of DIQKD. Especially, since these setups are entirely optical, they potentially allow for high repetition rates ($\approx 10^8\ Hz$ \cite{Gisin2010}).

The main drawback of this implementation, however, is that on-demand high fidelity single photon sources are currently out of reach. Heralded and imperfect single photon sources can always be used (\textit{e.g} with spontaneous parametric down conversion (SPDC) processes, using the signal mode as ``trigger''), but in this case the success probability of the QND measurement decays approximately to $\eta_d^2 (1-T)p^2 \approx 10^{-8}$ \footnote{$\eta_d$ is the trigger detectors efficiency and $p$ is the probability to produce a singlet in the SPDC process. For details about this success probability, we refer the reader to \cite{Gisin2010}.}. Such success probability not only strongly diminishes the rate, but it also falls below the dark count rate of photo-detectors. Furthermore, imperfections in the single photon sources (\textit{e.g.} multiple pair emissions in SPDC) contribute as false-positives to the Bell state measurement, which in turn decrease the fidelity of the output mode and increase the efficiency threshold necessary to guarantee security.

A similar proposal by Curty and Moroder managed to slightly enhance the rates and decrease the critical efficiencies of the qubit amplifier \cite{Curty2011}. Such proposal is based on standard quantum relays for entanglement swapping with linear optics, but for the previously mentionned reasons, it also represents a great challenge.

A different alternative to implement DIQKD with a QND measurement following the architecture of \figref{fig.2} was  proposed in \cite{Mattar2013} with light-matter interaction systems, see also~\cite{Brunner2013}. In these schemes, the entanglement between photon pairs is transferred to solid state (spin) qubits mediated by cavity QED interactions. As this transfer is achieved in a heralded way, the spin is only measured when the transfer succeeds. The main advantage of such light-matter interaction schemes is that the spin state can subsequently be measured with near unit efficiency, the drawback being the fact that such spin read-out measurements take in general long times, which considerably limit the attainable repetition rates. Also, inefficient frequency conversion processes might be needed in order to successfully transfer the photonic state of the incoming system into the cavity. Note that very recently, such heralding mapping of a photonic qubit into an atomic state within a cavity has been experimentally achieved with a $\approx 90\%$ fidelity \cite{Norbert2015}.

\section{Post-selection on the outcome of a third party}
\label{sec.5}
\begin{figure}[h!]
  \begin{minipage}[c]{0.5\textwidth}
    \includegraphics[width=1.3\textwidth]{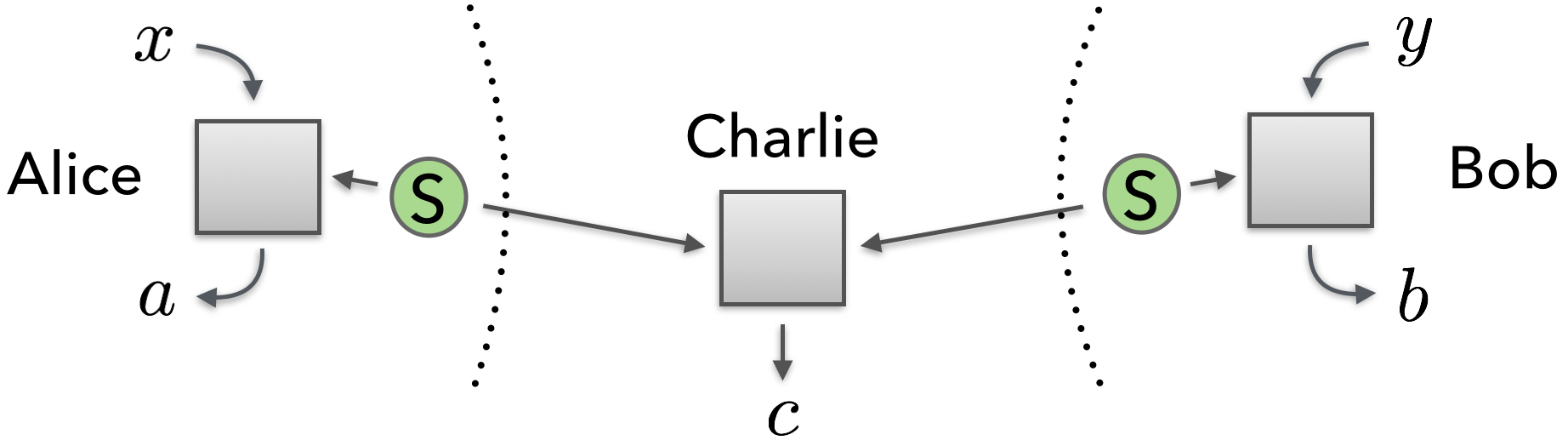}
  \end{minipage}\hfill
  \begin{minipage}[c]{0.5\textwidth}
    \caption{
       \textbf{DIQKD with a third party.} Depending on the outcome $c$, rounds can be safely  discarded at the end of the protocol, without opening the detection loophole.
    } \label{fig.3}
  \end{minipage}
\end{figure}

A different approach for overcoming channel loss in Bell experiments was proposed by Bell himself \cite{Bell1987}. The key idea is to record an additional signal to indicate whether the required state was successfully shared between Alice and Bob. This idea has been developed in entanglement swapping protocols \cite{Zeilinger98} and is to some extent the basis for quantum repeaters technologies \cite{Lutkenhaus2014}. For concreteness, and without loss of generality, such additional signal can be seen as the outcome $c$ of an untrusted measurement performed by a third party, Charlie, located somewhere between Alice and Bob (see \figref{fig.3}). Crucially, the outcome $c$ has to be independent of the choices of measurements and outcomes made by Alice and Bob. In a standard Bell experiment, this can be guaranteed through space-like separation. In DIQKD, this independence is guaranteed by default as Alice and Bob have full control of all the information about $a$, $b$, $x$ and $y$ that exits their secure locations.

Hence, by conditioning the validity of rounds on the outcome $c$, failed distribution events (those for which Alice and Bob's particles sent to Charlie got lost in the channel) can be safely excluded at the end of the protocol. In other terms, because of the independence of $c$ from $x$, $y$, $a$ and $b$, the scheme depicted in \figref{fig.3} can be simply seen as a heralded preparation from Charlie to Alice and Bob, with the remarkable advantage that no channel loss occurs between the sources S and the main users.

This idea has been implemented in the recent years in several light-matter interaction systems, successfully closing the detection loophole \cite{Pironio2010,Weinfurter2012}, and even attaining a loophole-free realisation~\cite{Hansen2015}. Initially, matter-photon entanglement is created at each site. Subsequently, photons are sent to Charlie who performs a joint measurement that swaps the entanglement to the matter spins. Alice and Bob then read-out their spins with near unit efficiency. Finally, using classical communication, only rounds for which Charlie's measurement succeeded are kept.

As mentioned earlier, the principal inconvenient of working with matter systems is the slow times required to perform the read-out (tipically few $\mu s$). Naturally, this strongly limits the number of secret bits that can be certified per time unit. Thus, an interesting alternative would be to analyze the architecture for DIQKD of \figref{fig.3} within all-optical implementations, which potentially allow for much higher repetition rates. We will report on this matter elsewhere \cite{MattarUnpublished}.

\section{Discussion}
\label{sec.6}

We have presented and discussed two distinct architectures to overcome the difficult problem of channel loss in the frame of DIQKD physical implementations. The main drawback of the first method (local heralding \secref{sec.3}) with respect to the second one (post-selection with a third party \secref{sec.4}) is the necessity for an ancilla to perform the QND measurement. It is difficult to say if the future development of quantum memories will soon be sufficient to consider seriously the use of such ancillary systems, but for the time being, the second solution seems easier to achieve. In fact, as we mentioned earlier, only the second solution has successfully closed the detection loophole, and in several occasions \cite{Pironio2010,Weinfurter2012}, recently together with the locality loophole \cite{Hansen2015}.

It is also worth noting a general trade-off occurring in both solutions, between high detection efficiency and high repetition rate. The use of matter systems (spins) allows for high detection efficiencies, but the repetition rate is strongly limited by the long times required to prepare and read-out the spins. On the other hand, entirely optical systems benefit from high repetition rates as measurements are generally fast and no re-initialization of the system is required at each round, but the detection process in such optical systems is difficult due to coupling losses and inefficiency of photo-detectors. Near future development of either photo-detectors or faster fluorescent methods for matter systems could lean the balance in favor of either purely optical systems or matter-based ones.

\section{Acknowledgements}
This work is supported by the Mexican CONACYT graduate fellowship program, the ERC CoG QITBOX, the EU project SIQS, the John Templeton Foundation, the Spanish project FOQUS (FIS2013-46768-P), the Generalitat de Catalunya (SGR875) and the Quantum Cryptography AXA Chair for Data Protection.


\bibliographystyle{iopart-num}
\bibliography{biblio}

\end{document}